\documentclass[aip,jcp,preprint]{revtex4-1}
\usepackage{amsmath}
\usepackage{amssymb}
\usepackage{graphicx}
\usepackage[usenames,x11names,svgnames,dvipsnames]{xcolor}
\usepackage[colorlinks]{hyperref}
\usepackage{cancel}
\usepackage{enumerate}
\usepackage{braket}

\newcommand{\vb}[1]{{\boldsymbol {#1}}}
\newcommand{\dd}{\scriptstyle D}
\newcommand{\od}{\scriptstyle OD}
\begin{document}
\title  {
            Efficient simulation of wave-packet dynamics
            on multiple coupled potential surfaces
            with split potential propagation
        }

\author{  Avi Pe'er }
\email{avipeer@mail.biu.ac.il}
\author{ Igal Aharonovich }
\email{jigal2@gmail.com}
\affiliation{Department of Physics and BINA center for nano-technology, Bar-Ilan University, Ramat Gan 52900, Israel}
\date{\today}
\begin{abstract}
We present a simple method to expedite simulation of quantum wave-packet dynamics by more than a factor of $2$ with the Strang split-operator
propagation.
Dynamics of quantum wave-packets are often evaluated using the the \emph{Strang} split-step propagation,
where the kinetic part of the Hamiltonian $\hat{T}$ and the potential part $\hat{V}$
are piecewise integrated according to $e^{- i \hat{H} \delta t} \approx e^{- i \hat{V} \delta t/2} e^{- i \hat{T}\delta t} e^{- i \hat{V} \delta t/2}$,
which is accurate to second order in the propagation time $\delta t$.
In molecular quantum dynamics, the potential propagation occurs over multiple coupled potential surfaces
and requires matrix exponentiation for each position in space and time which is computationally demanding.
Our method employs further splitting of the potential matrix $\hat{V}$
into a diagonal space dependent part $\hat{V}_{D}(R)$
 and an off-diagonal time-dependent coupling-field  $\hat{V}_{OD}(t)$,
 which then requires only a single matrix exponentiation for each time-step,
considerably reducing the calculation time even in the simplest two-surface interaction
($\sim$70\% reduction observed in potential propagation time).
We analyze the additional error due to the potential splitting and show it to be small compared to the inherent error associated
with the kinetic/potential splitting.
\end{abstract}

\maketitle

\section{Introduction}
Molecules are considerably more complex than atoms.
The additional degrees of freedom of the interacting nuclei dramatically affect their
quantum dynamics % and interaction with external fields,
which are key to  understanding molecular phenomena, such as association, dissociation,
chemical reactions and interaction with coherent time-dependent fields
\cite{:/content/aip/journal/jcp/84/7/10.1063/1.450074,:/content/aip/journal/jcp/83/10/10.1063/1.449767,0034-4885-66-6-201}.
The quantum dynamics of molecules is generally treated with the Born-Oppenheimer approximation \cite{atkins2011molecular},
which exploits the large difference in mass between electrons and the nuclei,
to adiabatically separate the nuclear degrees of freedom from the electronic degrees of freedom.
The total state of the molecule can then be specified as
\begin{align}
    \label{eq:born-oppenheimer_decomposition}
    \psi_{\text{mol}}(r,R,t) & = \psi_{N}(R,t) \psi_{e}(R;r),
\end{align}
where $R$ ($r$) is the nuclear (electronic) configuration and electronic eigenstates $\psi_{e}(R;r)$ are evaluated with the configuration of the nuclei
fixed at $R$.
The electronic eigenenergies $V^{(e)}_{n}(R)$, which are parameterized by the nuclear configuration $R$,
serve as the potential for the nuclear dynamics, essentially dividing the nuclear
system between distinct potential surfaces.
On each surface, the nuclei evolve according to the Schr\"{o}dinger equation
\begin{align}
    \begin{split}
        \label{eq:schrodinger_equation_generic}
        i \hbar \dfrac{\partial}{\partial t}\psi_{n}(R,t)
        & =
            (\hat{T} + \hat{V}_{n}) \psi_{n}(R,t)
%         \\
%         & \quad
            - \sum_{m \neq n} \vb{\mu}_{mn} \cdot \vb{E}_{mn}(t) e^{i \omega_{mn} t} \psi_{m}(R,t),
    \end{split}
\end{align}
where $\psi_{n}(R,t)$ is the nuclear wave-function on the $n^{\text{th}}$ electronic surface,
$\hat{V}_{n}(R)$ is the electronic eigenenergy
and $\vb{E}_{mn}(t)$ is an external field which couples the electronic potential surfaces through the electronic dipole $\vb{\mu}_{mn}$.
Equation \eqref{eq:schrodinger_equation_generic} employs the \emph{dipole approximation}
(fields are homogeneous across the size of the molecule),
the \emph{rotating wave approximation}
and the \emph{Condon approximation} (electronic dipole $\vb{\mu}_{mn}$ is independent of the nuclear configuration).
One can express \eqref{eq:schrodinger_equation_generic} as the dynamic evolution of a multi-surface state vector
$\ket{\psi(R,t)}$ according to the matrix valued operator $\hat{H} = \hat{T} + \hat{V}$:
\begin{align}
    i \dfrac{\partial}{\partial t} \ket{\psi(R,t)} & = (\hat{T} + \hat{V}) \ket{\psi(R,t)}
\end{align}
where $\hat{V}(R,t)$ includes the potential surfaces and external fields in a multi-surface matrix of the form
\begin{align}
    \label{eq:U_matrix}
    \hat{V}(R,t)
    & =
        \begin{pmatrix}
            V_{1}(R) - \hbar \omega_1                                               &   - \vb{\mu}_{12} \cdot \vb{E}_{12}(t) e^{i \omega_{12} t}    & \ldots
            \\
            - \vb{\bar{\mu}}_{12} \cdot \vb{\bar{E}}_{12}(t) e^{- i \omega_{12} t}  &   V_{2}(R)  - \hbar \omega_2                  & \ldots
            \\
            \vdots
        \end{pmatrix}
\end{align}
In the diagonal terms $\hat{V}_{mm}(R,t) = V_{m}(R) - \hbar \omega_{m}$, $V_{m}(R)$ is the potential surface and  $\omega_{m}$
is the rotation frequency of the state $\psi_{m}(R)$.
The off-diagonal terms $\hat{V}_{mn}(R,t) = - \vb{\mu}_{mn} \cdot \vb{E}_{mn} e^{- i \omega_{mn}t} \equiv \Omega_{mn}(t) e^{- i \omega_{mn}t} $ are
the electronic dipole couplings between  different surfaces.
Evidently, the diagonal terms depend only on the spatial coordinates $R$, whereas the off-diagonal terms
depend, to a very good approximation, only on time $t$.

Of the many techniques to evaluate \eqref{eq:schrodinger_equation_generic} \cite{0034-4885-58-4-001},
we focus on the commonly used split-operator method \cite{marchuk_1968,strang_1968,decomposition_methods_for_differential_equations},
where the exact unitary evolution over a small time-step $\ket{\psi(t + \delta t)} = e^{-i \hat{H}(t) \delta t} \ket{\psi(t)}$
is approximated by a successive application of the constituent operators $e^{- i a \hat{T} \delta t} \ket{\psi(t)}$ and $e^{- i b \hat{V} \delta t} \ket{\psi(t)}$ where $a$ and $b$
depend on the particular splitting method.
Using the popular $A-B-A$ splitting given by Strang \cite{0034-4885-58-4-001,strang_1968}, we obtain
\begin{align}
    \label{eq:strang_split_step_operator}
    \ket{\psi(t + \delta t)} & %\approx e^{- \frac{i}{\hbar} \hat{H}(t) \delta t } \ket{\psi(t)}
                               \approx e^{- \frac{i}{\hbar} \hat{V}(t) \delta t / 2}
                                \cdot
                                 e^{- \frac{i}{\hbar} \hat{T}    \delta t    }
                                \cdot
                                 e^{- \frac{i}{\hbar} \hat{V}(t) \delta t / 2}
                                 \ket{\psi(t)}
                                 ,
\end{align}
which is accurate to second order in $\delta t$,
according to the well known relation \cite{decomposition_methods_for_differential_equations}
\begin{align}
    \label{eq:split_operator_with_1st_correction_term}
    e^{\delta t( \hat{A} + \hat{B} )}
    & =
        e^{\frac{\delta t}{2} \hat{A}}
        e^{t \hat{B}}
        e^{\frac{\delta t}{2} \hat{A}}
        -
        \dfrac{\delta t^3}{24} [[\hat{A},\hat{B}], \hat{A} + 2 \hat{B}]
        +
        O(\delta t^4)
\end{align}
where the error of order $\delta t^3$ is also shown.

A useful aspect of the \emph{kinetic/potential} splitting is the possibility
to propagate the kinetic part $e^{- \frac{i}{\hbar}\hat{T} \delta t}$ by means of a fast Fourier-transform.
% which has superior computational properties when compared to explicit differentiation \cite{0034-4885-58-4-001}.
The potential propagation $e^{- \frac{i}{\hbar} \hat{V}(R,t) \delta t}$ is
normally performed by explicit numerical diagonalization and exponentiation of the
multi-surface potential matrix $\hat{V}(R,t)$ at every point in the space-time $(R,t)$.
% Though this can be easily carried out analytically for two potential surfaces (as in \cite{0034-4885-58-4-001}),
% it becomes challenging in higher number of coupled potential surfaces.
% Furthermore, even though the kinetic Fourier step is considered the most computationally consuming part of the propagation,
% numerical exponentiation at every space-time point has a considerable cost as well.
% % % Even for the simplest two-surface case, we measured $\sim 50\%$ shorter overall execution time (kinetic and potential propagation)
% % % when using the split potential propagation.
% for the split potential propagation
% when compared to the standard method.

\section{Splitting the potential operator $\hat{V}(R,t)$}
We propose a simple optimization of the evaluation of $e^{- \frac{i}{\hbar} \hat{V}(R,t) \delta t}$,
which requires only a single matrix exponentiation for every time-step. % (for the entire space).
% considerably simplifying and expediting the computation.
By distinguishing the diagonal and off-diagonal terms in $\hat{V}(R,t)$ \eqref{eq:U_matrix},
\begin{align}
    \label{eq:potential_decomposition}
    \hat{V}(R,t) & = \hat{V}_{\dd}(R) + \hat{V}_{\od}(t),
\end{align}
$\hat{V}$ is split, to a very good approximation, into \emph{spatial} only $\hat{V}_{D}(R)$ and \emph{temporal} only $\hat{V}_{OD}(t)$.

Using the Strang splitting again, \eqref{eq:split_operator_with_1st_correction_term} we find
\begin{align}
    \begin{split}
    \label{eq:general_potential_splitting}
    e^{- \frac{i}{\hbar} \hat{V}(R,t) \delta t}
    & =
        e^{- \frac{i}{\hbar} \hat{V}_{OD}(t) \delta t/2  }
        e^{- \frac{i}{\hbar} \hat{V}_{D}(R) \delta t    }
        e^{- \frac{i}{\hbar} \hat{V}_{OD}(t) \delta t/2  }
%     \\
%     & \qquad
        +
        O(\delta t^3),
    \end{split}
\end{align}
where evaluation of the diagonal spatial term $e^{- \frac{i}{\hbar} \hat{V}_{D}(R) \delta t    }$ is trivial (exponent of a diagonal matrix),
and evaluation of the off-diagonal temporal term $e^{-\frac{i}{\hbar} \hat{V}_{OD}(t) \delta t}$ requires
diagonalization of $\hat{V}_{OD}(t)$, which now
% has a rank of $N-1$ (at least one of its eigenvalues is identically zero).
% Importantly, the  diagonalization of $\hat{V}_{OD}(t)$ is independent of the spatial coordinates $R$, and
can be evaluated only once every time step,
substantially reducing the computational cost.

Even for the simple, well-established two-surface propagation \cite{0034-4885-58-4-001},
the potential splitting brings considerable computational benefit.
Consider the two-surface potential \cite[eq.~2.46]{0034-4885-58-4-001}:
\begin{align}
    \label{eq:two_potential_matrix_form}
    \hat{V}(R,t)
    & =
        \begin{pmatrix}
            V_{1}(R)        &   \Omega(t)
            \\
            \bar{\Omega}(t) &   V_{2}(R) - \hbar \omega
        \end{pmatrix}
        ,
\end{align}
where $\Omega(t)$ is the \emph{coupling potential}
\begin{math}
    \Omega(t)
    =
        \mu_{12} E_{12}(t)
\end{math}.

The \emph{standard} two-surface potential propagator \cite[See][eq.~2.47]{0034-4885-58-4-001}
is then the exponentiation of \eqref{eq:two_potential_matrix_form}, which can be carried out analytically
\begin{align}
    \label{eq:standard_two_surface_potential_propagator}
    e^{- \frac{i}{\hbar} \hat{V}(t) \delta t}
    & =
        \exp
        \Big[
            -
            i
            \dfrac{\delta t}{2\hbar}
            ( V_{1}(R) + V_{2}(R) - \hbar \omega )
        \Big]
        \begin{pmatrix}
            A           &   B
            \\
            \bar{B}     &   \bar{A}
        \end{pmatrix}
        ,
\end{align}
where
\begin{align}
    \begin{split}
        A
        & =
            \cos(\phi)
            -
            i \dfrac{\Delta  t}{\hbar} \Delta V(R)
            \dfrac{\sin(\phi)}{\phi}
            ,
    \\
        B
        & =
            -
            i \dfrac{\delta t}{\hbar} \bar{\Omega}(t)
            \dfrac{\sin(\phi)}{\phi}
    \end{split}
\end{align}
and
\begin{align}
    \begin{split}
    \label{eq:delta_U_R}
        \Delta V(R)
        & =
            \dfrac{1}{2}
            \Big(
                V_{1}(R) - V_{2}(R) + \hbar \omega
            \Big)
        ,
    \\
        \phi(R,t)
        & =
            \dfrac{\delta t}{\hbar}
            \sqrt{ |\Omega(t)|^2 + \Delta V^2(R)}
    \end{split}
\end{align}

In our \emph{split potential propagator} method, we first decompose $\hat{V}(R,t)$ into
\begin{align}
    \begin{split}
    \label{eq:U_split_to_U_D_and_U_OD}
    \hat{V}(R,t)
    & =
        \hat{V}_{D}(R)
        +
        \hat{V}_{OD}(t)
    \\
    & =
        \begin{pmatrix}
            V_{1}(R)    &   0
            \\
            0           &   V_{2}(R) - \hbar \omega
        \end{pmatrix}
        +
        \begin{pmatrix}
            0                &   \Omega(t)
            \\
            \bar{\Omega}(t)  &   0
        \end{pmatrix}
    \end{split}
\end{align}
which leads to the \emph{split potential propagator}:
\begin{align}
    \label{eq:two_surface_split_propagator}
    \begin{split}
        & e^{- \frac{i}{\hbar} \hat{V}(t) \delta t}
        \approx
            e^{- \frac{i}{\hbar} \hat{V}_{OD} \delta t/2}
            \cdot
            e^{- \frac{i}{\hbar} \hat{V}_{D}(t) \delta t}
            \cdot
            e^{- \frac{i}{\hbar} \hat{V}_{OD} \delta t/2}
        \\
        & =
            \exp
            \Big[
                -
                i
                \dfrac{\delta t}{2\hbar}
                ( V_{1}(R) + V_{2}(R) - \hbar \omega )
            \Big]
            \times
        \\
        &   \qquad \qquad
            \times
            \begin{pmatrix}
                D(R) \cos(\alpha)                     &   - i \sin(\alpha) e^{i \theta}
                \\
                \\
                - i \sin(\alpha) e^{- i \theta}       &   \dfrac{1}{D(R)} \cos(\alpha)
            \end{pmatrix}
    \end{split}
\end{align}
where
\begin{align}
    \label{eq:alpha_and_theta_functions}
    \alpha(t)
    & =
        |\Omega(t)| \dfrac{\delta t}{\hbar}
    ,
        &
        \theta(t)
        & =
            \arg \Omega(t)
%         =
%             \tan^{-1}
%             \Big[
%                 \dfrac{ \text{Im } \Omega(t) }
%                       { \text{Re } \Omega(t) }
%             \Big]
\end{align}
and
\begin{align}
    \begin{split}
    D(R)
    & =
        \exp\Big[ - \dfrac{i \delta t}{\hbar} \Delta V(R) \Big]
    \\
    &
    =
        \exp
        \Big[
            -
            \dfrac{i \delta t}{2\hbar}
            \Big(
                V_{1}(R) - V_{2}(R) + \hbar \omega
            \Big)
        \Big]
    \end{split}
\end{align}

In the standard propagator form \eqref{eq:standard_two_surface_potential_propagator}, elements $A$ and $B$ depend non-trivially on $\phi(R,t)$,
and need to be evaluated individually for each point $R$ and time $t$.
In the split propagator form \eqref{eq:two_surface_split_propagator},
% on other hand, only the diagonal term $D(R)$ is space dependent,
% whereas the off-diagonal terms $\alpha(t)$ and $\theta(t)$ are time dependent,
% and therefore,
$\cos(\alpha), \sin(\alpha)$ and $e^{i \theta}$ need to be evaluated only once for each time step,
and $D(R)$ and the common phase
\begin{math}
    \exp
    (
        -
        i
        \frac{\delta t}{2\hbar}
        ( V_{1}(R) + V_{2}(R) - \hbar \omega )
    )
\end{math}
can be pre-computed once for the entire numerical execution.
In this way, the point-by-point spatial propagation becomes a simple $2\times 2$ matrix multiplication
with all of its elements pre-evaluated.

In performance studies with a two-surface propagation, the split potential method had an \emph{overall} 50\% reduction in calculation time,
including both the potential and the kinetic propagation (Fourier transforms).
When considering the potential propagation alone, the potential propagation time was cut by
$\sim 70\%$ compared to the  standard propagation. Details of the performance study are described in the appendix \ref{sec:details_on_numerical_study}.

\subsection{Error estimate}
Inherently, the secondary splitting of $\hat{V}(R,t)$ \eqref{eq:potential_decomposition} introduces only error of order $O(\delta t^3)$, which
is on par with the already existing kinetic/potential splitting \eqref{eq:strang_split_step_operator}.
However, even within the same order of accuracy, an estimate of the additional error is important to verify the applicability
of simulation parameters, such as the required time step for a desired calculation precision.
Thus, we provide in the following a simple semi-rigorous estimation of the additional error
(for more detailed and rigorous analysis, cf. \cite{decomposition_methods_for_differential_equations,tobias_2000}).

\begin{figure}[htb]
    \includegraphics[width=\linewidth]{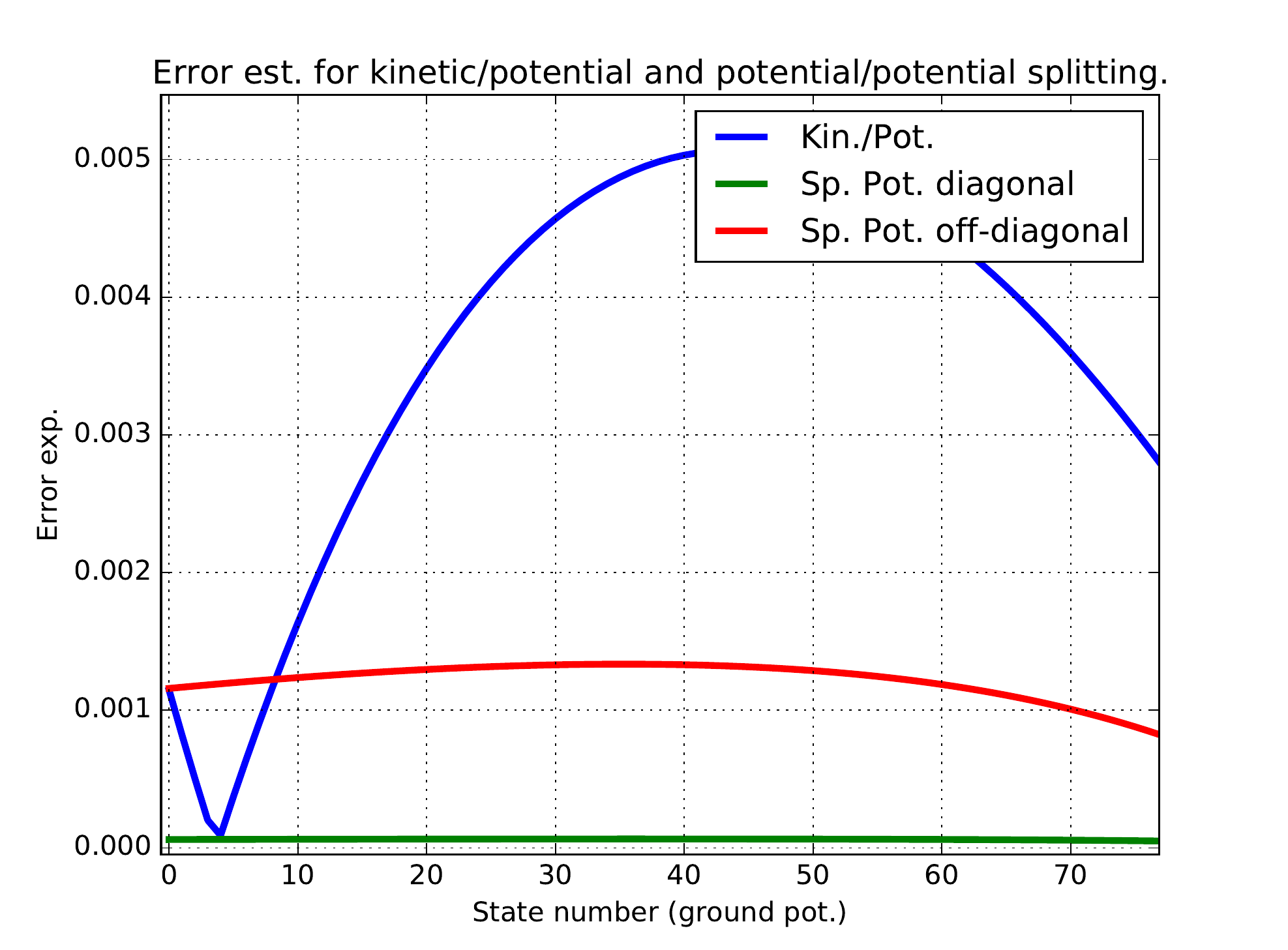}
    \caption{Comparison of the $3^{\text{rd}}$ order error estimate ${C}^{(3)}_{\text{kp}}$ \eqref{eq:kinetic_potential_splitting_3rd_order_correction}
             of the standard kinetic/potential splitting (blue)
             and the potential splitting $\hat{C}^{(3)}_{\text{pp}}$ \eqref{eq:potential_potential_splitting_3rd_order_correction}
             (red shows the off-diagonal part),
             plotted against the ground state eigenmode $\nu$.
             The value is dimensionless, and for unitary propagation, should be compared with unity.
             }
%              where $\hat{C}_{3}$ is the corresponding third order correction \eqref{eq:split_operator_with_1st_correction_term},
%              plotted against $\nu$, which numbers the ground potential modes of $K_2$.}
    \label{fig:error_estimate}
\end{figure}

We begin by evaluating the leading error term in \eqref{eq:split_operator_with_1st_correction_term},
of order $\delta t^3$,
for both the kinetic/potential splitting and the potential/potential splitting.
The leading error term of $e^{a(A+B)} \approx e^{aA/2} e^{aB} e^{aA/2} + \hat{C}^{3} + \ldots $ is given by
\begin{align}
    \label{eq:third_order_correction}
    \hat{C}^{3}
    & =    a^3
           [[\hat{A},\hat{B}], \hat{A} + 2\hat{B}]
\end{align}
In order to gain insight on its effect, we calculate the expectation value of $\bra{\nu} \hat{C}^{3} \ket{\nu}$ for the kinetic/potential
and the potential/potential splitting for a given complete basis $\ket{\nu}$, which we chose as the complete set of bound eigenstates
of the ground potential.
For the kinetic/potential leading  error term is
\begin{align}
    \label{eq:kinetic_potential_splitting_3rd_order_correction}
    C^{(3)}_{\text{kp},\nu}
    \equiv
    \bra{\nu} \hat{C}^{(3)}_{\text{kp}} \ket{\nu}
    & =     \dfrac{\delta t^3}{24 \hbar^3}
            \bra{\nu}
                [[\hat{T}, \hat{V}], \hat{T} + 2 \hat{V}]
            \ket{\nu}
            .
\end{align}
Figure \ref{fig:error_estimate} shows $C^{(3)}_{kp,\nu}$
for all vibrational eigenmodes $\ket{\nu}$ of the ground potential (blue).
(For calculation of the eigenmodes $\ket{\nu}$ we used the Morse potential fits for the ground potential $X ^{1}\Sigma_{g}$ of Potassium dimer $K_2$ \cite{PhysRevA.54.204}).
Note that in this case, $\hat{C}^{(3)}_{\text{kp}} $ is a differential operator.

For the split potential propagation \eqref{eq:U_split_to_U_D_and_U_OD}, the corresponding $\hat{C}^{3}_{\text{pp}}$
error estimate for a coupled two surfaces is
\begin{align}
    \label{eq:potential_potential_splitting_3rd_order_correction}
    \begin{split}
        C^{(3)}_{\text{pp},\nu \mu}
        & \equiv
            \bra{\nu^{(e)}, \nu^{(g)}} \hat{C}^{(3)}_{\text{pp}} \ket{\nu^{(e)}, \nu^{(g)}}
        \\
        & =     \dfrac{\delta t^3}{24 \hbar^3}
%         \\
%         & \quad
%             \times
                \bra{\nu^{(e)}, \nu^{(g)}}
                    [[\hat{U}_{\dd}, \hat{U}_{\od}], \hat{U}_{\dd} + 2 \hat{U}_{\od}]
                \ket{\nu^{(e)}, \nu^{(g)}}
                ,
    \end{split}
\end{align}
where $\nu^{(e)} ( \nu^{(g)})$ are the corresponding eigenmodes in the excited (ground) potential surfaces, respectively.
In the two-surface case \eqref{eq:two_potential_matrix_form}, $\hat{C}^{(3)}_{pp}$ takes the form
\begin{align}
    \label{eq:two_surface_split_potential_propagator_correct_term_full}
    \begin{split}
        \hat{C}^{(3)}_{\text{pp}}
        & =
            -
            |\Omega(t)|
            \Delta V(R)
            \begin{pmatrix}
                4 |\Omega(t)|                       &   e^{i\theta(t)} \Delta V(R)
                \\
                e^{- i \theta(t)} \Delta V(R)       &   -4|\Omega(t)|
            \end{pmatrix}
    \end{split}
    .
\end{align}
% where $\Delta V(R)$ is given in \eqref{eq:delta_U_R} and $\theta(t) = \text{arg} [ \Omega(t)] $ \eqref{eq:alpha_and_theta_functions}.

Taking the transition from ground $X ^{1}\Sigma_{g}$ to excited $A ^{1} \Sigma_{u}$ of $K_2$ %potassium dimer molecules as the ground/excited potential surfaces,
the expectation value of the off-diagonal term  %$\hat{C}^{(3)}_{\text{pp}}$
\begin{math}
    \bra{\nu^{(g)}} |\Omega(t)| \Delta^2 V(R) \ket{\nu^{(g)}}
\end{math}
is shown in red in figure \ref{fig:error_estimate}, for all $\nu$.
Strictly speaking, the off-diagonal terms couple only ground/excited wave-functions, yet its application
on ground/ground states is also useful to estimate its norm.
Similarly, the green curve shows  the diagonal part
\begin{math}
    \bra{\nu^{(g)}} |\Omega(t)|^2 \Delta V(R) \ket{\nu^{(g)}}
\end{math}
,
which is considerably smaller.
The corresponding plots for the excited potential are not shown, since they have very similar values.
It is clear from figure \ref{fig:error_estimate} that the additional error for the split-potential is not larger
(and normally much smaller) than that of the kinetic/potential splitting.
Note that we assumed for this error estimation, an extremely high coupling field  $\Omega \sim 1.5\text{THz}$ (Rabi freq.)
that would occur only with very intense optical pulses.
$\Delta V(R)$, on the other hand, depends on the difference in the position of the potential dip,
and can be roughly estimated as half of the potential depth $D_{e/g}$ for either the excited or ground potential curves.
For instance, the potential depth of the surface  $A ^{1} \Sigma_{u}$ of $K_2$ is $D_e \sim 6300 \text{cm}^{-1} \sim 190 \text{THz} \gg |\Omega(t)|$.

The split potential error term $\hat{C}^{(3)}_{\text{pp}}$ is inherently different from
the kinetic/potential $\hat{C}^{(3)}_{\text{kp}}$. The kinetic/potential error depends on the steepness of the potential for each potential curve,
whereas $\hat{C}^{(3)}_{\text{pp}}$ depend on the overall landscape variation of the potential curves, as well as on the coupling field $|\Omega(t)|$.
Also, from \eqref{eq:two_surface_split_potential_propagator_correct_term_full},
it is clear that the off-diagonal terms are linear in the coupling potential $\Omega(t)$ and quadratic in the difference between the potentials $\Delta V(R)$,
(and vice-versa for the off-diagonal terms).

To obtain a \emph{global error} estimation,
we also  tested our analysis by numerical calculation of the dynamics of a particular $K_2$ molecule for a prescribed intense field
and compared the results of the standard and the split potential methods,
using various time-steps and field strengths.
In particular, we computed the excited wave-packet $\psi_{e}(t)$ evolving under a coupling field comprising of a train of 40 identical Gaussian pulses (15fs FHWM) spanning nearly $50$ps.
We compared the simulation results for various peak powers and time steps.
Moreover, we considered the solution $\ket{\psi_e^{(0.1)}}$ of the standard propagation with smallest time-step $dt = 0.1\text{fs}$ as the \emph{reference solution},
and compared to it all the results of longer time-steps $\ket{\psi_e^{(dt)}}$ by calculating the overlap
$K_{n,i} = |\braket{ \psi_e^{(0.1)}|\psi_e^{(dt)}}|$.
The deviation of $K_{n,i} \leq 1$ from unity  is taken as a benchmark for estimation of the calculation error.

\begin{figure}
    \includegraphics[width=\linewidth]{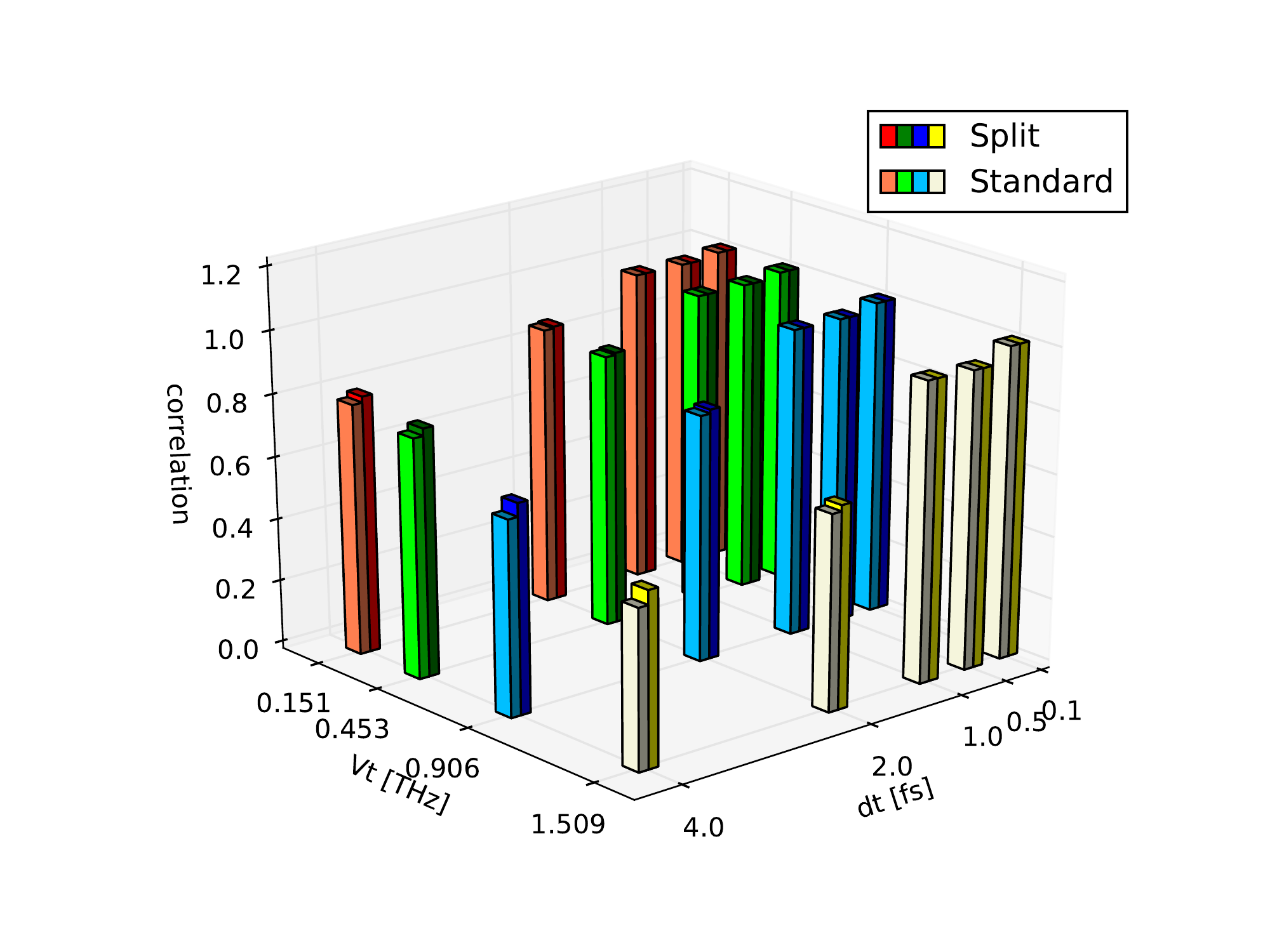}
    \caption{Overlap of the excited potential wave packet at the end of a simulation with a
             \emph{nominal} very small step-size $dt = 0.1\text{ fs}$ taken as a reference solution,
             both for standard and split potential propagators, for various steps sizes and field strengths.
             Each case simulated identical set of strong pulses (with a max amplitude $V_n$[THz] shown in the plot).
             The split and standard propagators show nearly identical correlation properties for all time-steps and field strengths,
             and both diminish for a longer step size and stronger coupling field.
             Value of $1.0$ for the overlap indicates perfect agreement with the reference solution.
             }
    \label{fig:correlation}
\end{figure}

$K_{n,i}$ is plotted in figure \ref{fig:correlation}, where for every pair $(dt_i, V_n)$,
a double bar denotes the overlap value for both standard propagation and split potential propagation methods.
Evidently, both methods show similar
overlap qualities for the same time-step and both diminish nearly identically for increased coupling strength and for longer time-step.
Therefore, up to reasonably strong fields, the potential splitting does not add any noticeable error beyond the already existing error
of the kinetic/potential splitting.
The $50\%$ improvement in overall execution time observed for the two-surface system ($\sim70\%$ for the potential part alone)
now leaves most of the
computational burden on the kinetic Fourier propagation
(see appendix \ref{sec:details_on_numerical_study}).

\section{Conclusions}
Split potential propagation offers a significant computational benefit over direct exponentiation of the potential matrix
with no significant additional error compared to the standard kinetic/potential splitting.
Even for the simple two-surface case, $50\%$ reduction in overall execution time was measured compared to the standard potential propagation.
The benefit is expected to be even more substantial for higher dimensional couplings of $3$ or more surfaces,
as exponentiation of higher order matrices is much more demanding than the simple $2\times 2$.
Thus, since the off-diagonal coupling matrix is of lower rank  ($n-1$ instead of $n$) and since its exponentiation is needed
only once per time-step, it considerably shortens the evaluation time of the potential propagation,
which is the most time-consuming part of the entire propagation.
%
% Splitting the potential does introduce a new source of error \eqref{eq:potential_potential_splitting_3rd_order_correction},
% which depends on both the coupling field $\Omega(t)$ and the difference in potential landscape $\Delta V(R)$ \eqref{eq:two_surface_split_potential_propagator_correct_term_full},
% but this error is on par with the already existing kinetic/potential splitting,
% and scales linearly with the field.
% It was demonstrated to contribute no significant errors for reasonably strong fields.

Higher order precision can be achieved using higher order splittings
by multiple evaluation of $e^{- i a_i\hat{V}_{D} \delta t}$ and $e^{- i b_j \hat{V}_{OD} \delta t}$ for various values of $a_i$ and $b_j$
\cite[See sec. 3.1.5]{decomposition_methods_for_differential_equations},
which would also benefit from temporal/spatial splitting.

The method published here was used extensively in a recently published study of dynamics in a coherent Raman oscillator \cite{PhysRevLett.116.073603}.

This research was supported by the Israel Science Foundation (grants \#807/09 and \#46/14).

\clearpage
\appendix
\section{The numerical performance evaluation\label{sec:details_on_numerical_study}}
The $2 \times 2$ potential propagation was implemented in C++, compiled with GCC 5.2.1 and tested on Ubuntu Linux
running on a standard Core-i7 desktop with 8GB RAM.
When evaluating the wave-function at the desired final time, one generally coalesces
successive half-step potential propagation as
\begin{align}
    \label{eq:half_step_to_full_step}
    \begin{split}
    \psi(t + T)
    & =
        \prod_{j=0}^{T/\delta t}
            e^{- i \hat{V} \delta t/2}
            e^{- i \hat{T} \delta t  }
            e^{- i \hat{V} \delta t/2}
            \cdot
            e^{- i \hat{V} \delta t/2}
            \ldots
            e^{- i \hat{V} \delta t/2}
    \\
      &  =
        \prod_{j=0}^{T/\delta t}
            e^{- i \hat{V} \delta t/2}
            e^{- i \hat{T} \delta t  }
            e^{- i \hat{V} \delta t  }
            e^{- i \hat{T} \delta t  }
            \ldots
            e^{- i \hat{V} \delta t/2}
            .
    \end{split}
\end{align}
In our study, we measured the execution time for a calculation of the state functions $\psi_{g}(R,t)$ and $\psi_{e}(R,t)$
using once the standard potential propagation \eqref{eq:standard_two_surface_potential_propagator} and then the split propagation \eqref{eq:two_surface_split_propagator}.
As a test case we took the $K_2$ molecular dimer simulated over $10^{6}$ time-steps on a 512 point spatial grid.
The kinetic propagation
\begin{math}
    e^{- i \hat{T} \delta t} \psi(R,t)
    =
        \mathcal{F}^{-1} e^{ - i \frac{k^2}{2m} \delta t } \mathcal{F} \psi(R,t)
        ,
\end{math}
which requires two Fourier transforms for each potential surface,
was implemented using the well-established FFTW library \cite{1386650}.

The results are described in table \ref{table:simulation_results_comparison}, where the added cost of evaluating transcendental functions $\cos$ and $\sin$
for each grid point and time-step are evident.
\begin{table}[b]
    \begin{tabular}{ p{3.5cm} p{2cm} p{2cm} p{4cm} }
         \hline
         \hline
             Propagation type   &   Method  &   Count     &   Duration (seconds)
             \\
         \hline
         \hline
             Potential          &   Standard&   $10^{6}$  &   35
             \\
                                &   Split   &   $10^{6}$  &   11
             \\
             Kinetic            &   -       &   $10^{6}$  &   15
             \\
         \hline
    \end{tabular}
    \caption{Comparison of the standard and split potential propagation along with the kinetic propagation
             for $10^{6}$ time-steps. The system simulated was a $K_2$ dimer with 512 spatial grid points,}
    \label{table:simulation_results_comparison}
\end{table}

\clearpage
%%\bibliography{../jigal_university}
\bibliography{sts}

\begin{thebibliography}{10}

\bibitem{:/content/aip/journal/jcp/84/7/10.1063/1.450074}
Moshe Shapiro and Paul Brumer.
\newblock {Laser control of product quantum state populations in unimolecular
  reactions}.
\newblock {\em The Journal of Chemical Physics}, 84(7):4103--4104, 1986.

\bibitem{:/content/aip/journal/jcp/83/10/10.1063/1.449767}
David~J. Tannor and Stuart~A. Rice.
\newblock {Control of selectivity of chemical reaction via control of wave
  packet evolution}.
\newblock {\em The Journal of Chemical Physics}, 83(10):5013--5018, 1985.

\bibitem{0034-4885-66-6-201}
Moshe Shapiro and Paul Brumer.
\newblock {Coherent control of molecular dynamics}.
\newblock {\em Reports on Progress in Physics}, 66(6):859, 2003.

\bibitem{atkins2011molecular}
P.W. Atkins and R.S. Friedman.
\newblock {\em {Molecular Quantum Mechanics}}.
\newblock OUP Oxford, 2011.

\bibitem{0034-4885-58-4-001}
B~M Garraway and K~A Suominen.
\newblock {Wave-packet dynamics: new physics and chemistry in femto-time}.
\newblock {\em Reports on Progress in Physics}, 58(4):365, 1995.

\bibitem{marchuk_1968}
G.~I. Marchuk.
\newblock {Some Applications of Splitting-up Methods to the Solution of
  Mathematical Physics Problems}.
\newblock {\em Aplikace Matematiky}, 13(2):103--132, 1968.

\bibitem{strang_1968}
Gilbert Strang.
\newblock {On the Construction and Comparison of Difference Schemes}.
\newblock {\em SIAM J. Numer. Anal.}, 5(3):506--517, September 1968.

\bibitem{decomposition_methods_for_differential_equations}
Juergen Geiser.
\newblock {\em {Decomposition Methods for Differential Equations}}.
\newblock {Numerical Analysis and Scientific Computing}. CRC Press, 2009.

\bibitem{tobias_2000}
Christian~Lubich Tobias~Janhke.
\newblock {Error Bounds for Exponential Operator Splittings}.
\newblock {\em BIT Numerical Mathematics}, 44:735--744, 2000.

\bibitem{PhysRevA.54.204}
S.~Magnier and Ph. Milli{\'e}.
\newblock {Potential curves for the ground and numerous highly excited
  electronic states of ${\mathrm{K}}_{2}$ and NaK}.
\newblock {\em Phys. Rev. A}, 54:204--218, Jul 1996.

\bibitem{PhysRevLett.116.073603}
Igal Aharonovich and Avi Pe'er.
\newblock Coherent amplification of ultrafast molecular dynamics in an optical
  oscillator.
\newblock {\em Phys. Rev. Lett.}, 116:073603, Feb 2016.

\bibitem{1386650}
M.~Frigo and S.~G. Johnson.
\newblock The design and implementation of {FFTW3}.
\newblock {\em Proceedings of the IEEE}, 93(2):216--231, Feb 2005.

\end{thebibliography}
\bibliographystyle{unsrt}

\end{document}